# PV Power Forecasting Using Weighted Features for Enhanced Ensemble Method

Mohamed Massaoudi[1,2], Ines Chihi[3], Lilia Sidhom[3], Mohamed Trabelsi[2,4], Shady S. Refaat[2], Fakhreddine S. Oueslati[1]

[1]Unité de Recherche de Physique des Semi-Conducteurs et Capteurs, Carthage University, Tunis, Tunisia
[2]Department of Electrical and Computer Engineering, Texas A&M University at Qatar, Doha, Qatar
[3]LAboratory of Research in Automation (LA.R.A), National Engineering School of Tunisia, Tunisia
[4]Department of electronics and communications engineering, Kuwait College of Science, Kuwait
{mohamed.massaoudi, shady.khalil}@qatar.tamu.edu;{Ines.Chihi, Lilia.Sidhom}@enib.rnu.tn;
m.trabelsi@kcst.edu.kw; fakhreddine.oueslati@fst.rnu.tn

*Abstract*— Solar power becomes one of the most promising renewable energy resources in recent years. However, the weather is continuously changing, and this causes a discontinuity of energy generation. PV Power forecasting is a suitable solution to handle sudden disjointedness on energy generation by providing fast dispatching to grid electricity. These methods present a key insight into matchmaking grid electricity and photovoltaic plants. Bootstrap aggregation Ensemble method (Bagging) is classified as one of the most useful machine learning models which are applicable on supervised learning regression tasks. Following this regard, this paper proposes a state-of-art method based on bagging and this method works perfectly for PV power forecasting. The latter had powerful capabilities of tracking the behavior of stochastic problems with good accuracy with the aid of feature importance information. This approach comes to optimize bias/variance using feature weighting vector. Thus, this paper is devoted to present various feature importance techniques for Photovoltaic forecasting parameters. This technique consists of improving the aforementioned Ensemble model via contributing the knowledge expertise obtained from features analysis to be directly transformed into the Ensemble model. The proposed model is tested on PV power prediction. Therefore, the benchmarked technique shows an improvement in accuracy in terms of RMSE to 5%.

## I. INTRODUCTION

Nowadays, there is a huge demand for renewable energy sources (RES) in the modern economy due to industrial expansion and technological development[1]. The major advantage of RES is the natural clean replenishment in a short period of time by the existing flows of energy, from on-going natural processes[2]. These natural processes include sunshine, geothermal heat flows, wind, flowing water and biological processes. Among all renewable energy sources, solar energy witnesses major attention due to its advantages such as inexhaustibility and freedom from geographical restrictions. However, the discontinuity of solar energy generation causes a serious issue on the reliability of photovoltaic generation. This fact should be analyzed carefully to diminish their effect on grid stability. Unexpected weather parameters behavior threatens unit commitment and unbalance the demand/supply relationship. Baring this in mind, PV power forecasting is a key factor that decreases the impact of irradiation variability. In this respect, time series prediction is taking a lot of attention due to its importance in various applications. Forecasting is done through numerical patterns analysis between input parameters and the forecasted feature. These algorithms estimate the next photovoltaic energy. Therefore, four forecasting horizons are depicted: Very short term, short term, medium-term, and long-term predictions. In that sense, statisticians and scientists aim to provide an accurate forecast with less computational time and complexity. Hence, the precision is improved during many years to reach a minimum error with a reliable result. The aforementioned approaches for PV power forecasting are classified from physical and statistical methods machine learning (ML) algorithms. Physical models consist of transforming natural conversion equations into indicators of future behavior. These deal with linear systems and target short term predictions [3]. In that vein, solar power could be predicted using statistical methods such as autoregressive moving average ARMA and nonlinear autoregressive with exogenous values NARX[4]-[5]. These models are based on stochastic time series analysis[6]. There have proved their utility for short term forecasting and known by their simplicity since they did not use a large database for training or high computational hardware[7]. ML models are frequently used in prediction due to their high efficiency in tracking the predicted parameter[8]. Deep ML models are categorized into different classes including neural networks and reinforcement learning [9][10]. The latter had more attention in research and development[11]. artificial neural networks are used due to the non-linearity of weather parameters. Moreover, hierarchical forecasting presented in ensemble methods is highly effective on time series forecasting[12]. Boosting and bagging approaches are validated as reliable models of accurate forecasts in many tasks[13]-[14]-[15]. These forecasting methods use univariate or multivariate analysis[16], and they focus on direct or indirect photovoltaic forecasting by predicting a key element leading to an accurate PV power[17]-[18].

Assembling models in one box is an effective way to gain more accuracy than a single estimator. The aforementioned ensemble model has been given exponential importance during the last years. On the other hand, importance aided neural network (IANN) proposed by Ridwan in 2011 improves the accuracy of the neural network through integrating weighted metrics in features inputs[19]. The weights values are adjusted according to the importance of the input related to the desired output. The bigger the value, the closer the parameter to the predicted output. This method is used in various domains such as solar power prediction. Nevertheless, the weights may change from a feature importance method to another. Thus, in this study, the major contributions lie in:

1. Feature engineering analysis is done based on the weather dataset and using different approaches to analyze the behavior of system parameters.
2. The features are classified according to their weights to make a feature ranking upon the domain knowledge contribution.
3. A new approach based on multimodal is proposed. The weighted metric is applied to enhance the Ensemble model and increase accuracy.
4. The evaluation of the novel approach is made through a fair comparison with benchmarked algorithms.

Within this framework, this paper is divided as follows: First, a discussion about the ensemble methods used for time series forecasting is presented. Then, the proposed method is analyzed and tested based on Australian weather conditions Finally, some concluding remarks are mentioned.

## II. RELATED WORK

Photovoltaic power forecasting methods are classified according to different categories. The forecast horizon is one of the important factors for this classification. These lead to divide the approaches into short, medium and long-term prediction[20],[21]. The predicted power remains accurate for a number of timestamps ahead. Metric scores such as rooted mean square error or squared learning rate give a clear vision about the effectiveness of the method. Every class aims to target a specific item in the electricity grid. Usually, the longer the time span, the larger the error becomes, and the error is becoming bigger. Short forecasting horizon remains from seconds to maximum time steps of one day. This type of prediction is used to load dispatching and power quality. Nevertheless, medium-term forecasting spans from one day to one month and applied for maintenance planning[22]-[23], economical dispatch and grid stability. In addition, long term PV power forecasting is valid from one or two months to one year in order to fix a clear project planning and investment costs and benefits. From that standpoint, statistical models are often applied in short term predictions through statistical and probabilistic rules. The said models through these rules examines time series patterns to predict future energy. The analysis can be done either using univariate or multivariate time series forecasting. This provides a good accuracy with less computational time. On the other side, artificial neural networks (ANN) are very effective in handling medium- and long-term forecasting. Non-linearity of weather parameters made these approaches widely used, and outperform statistical methods in the majority of stochastic targets. ANN contains perceptron, input layers, hidden layers, and output layers. The layers are interconnected according to specific weights and bias of these perceptions. These parameters are fed to activation functions to make the propagation to the next layer. The output of the system can be written through Eq. (1).

$$U_N = b + \sum_{j=1}^{N}(W_j I_j) \quad (1)$$

With
- $U_n$   Final network output
- $b$   Bias
- $N$   Input number
- $W$   Weight connection
- $I$   Network input

Neural networks are frequently integrated into hybrid models. The goal here is to combine single models into one efficient system. A hybrid model usually increases the prediction accuracy. Therefore, there is an exponential focus on this type of mixture in research and development[24]–[26]. This theory is similar to another approach called the ensemble method. Nevertheless, there is a tiny difference between them. Ensemble methods are mainly used weak predictor to build a single strong learner. The said learners are homogenous and ideally Decision Trees[27]. The processing mechanism is carried out usually through bagging[28], boosting[29], stacking[27]. Then, the output can be taken through voting or averaging between the predictors. On the other side, the hybrid model uses heterogeneous algorithms approaches to build one predictive model[30]–[31]. These two methods gathered a lot of attention over years. High dimensional tasks require the said models to have the desired accuracy. Both of these techniques are used in PV power forecasting through many approaches such as SARIMA-RVFL [27] and GASVM techniques[27].

## III. STATEMENT AND CONTRIBUTION

Feature importance is given a lot of attention in classification and regression tasks. It consists of evaluating the features in relation to the output. This mission is done simply by removing one feature and calculating the prediction error without it. Then repeat the same process for the other parameters. This procedure allows the system to determine who much the features are informative for estimating the target behavior. Then, the non-important inputs may have to be removed to let only the mandatory parameters. it reduces the system complexity and computational time. The last step is assessing the accuracy of the model with the selected features to see the variance variability through cross-validation. This step is crucial to every prediction purpose. It gives a clear idea about the system parameters and thus choosing only the necessary parameters that majorly contribute to the prediction part. Baring this in mind, further information about the variance of the parameter is improving the model accuracy[19]. Importance aided neural network (IANN) is using this knowledge to optimize the variance/bias of the neural network. Future relative importance (FRI) introduces metrics weights to transport the important values to the model. In this paper, a novel Voted Feature Weighting (VFW) is introduced. The weights are fed in an ensemble learning system for the aim of getting further accuracy. The Australian weather is deeply analyzed from many parameters and the medium-term forecasting is assessed.

## IV. BIAS-VARIANCE TRADE-OFF

Classifiers and regressors' accuracy are related essentially to three elements: bais, variance, and noise. The bad value of these items is leading either to overfitting or underfitting. A better understanding of these elements is a key factor to improve prediction results. Typically, ML bais is the mismatched measurement values between reality and predictions taken from the learning phase. i.e. How much the prediction is distant from the ground truth. While the variance is the estimation of the squared deviation of an aleatory feature from its mean[32], [33]. Therefore, errors occur from high variance or high bais. To simplify mathematically the items, given $Y$ as a feature output predicted through a function $f$ with a set of inputs $X$. assuming that $\hat{f}$ is an estimation of $f(x)$. Then, the definition of error is presented in Eq. (2):

$$Err(x) = E[(Y - \hat{f}(x))^2] \quad (2)$$

With an output $Y$ written as following in the Eq. (3).

$$Y = f(X) + \varepsilon \quad (3)$$

Thus, the error can be written in Eq. (4).

$$Err(x) = (E[f(x)] - f(x))^2 + E[(f(x) - E[f(x)])^2] + \sigma_\varepsilon^2 \quad (4)$$
$$= Bias^2 \quad + Variance \quad + Noise$$

The ideal target is minimizing both variance and bais at the same time to reduce the errors. However, the latter elements are inversely proportional which poses a serious dilemma on how to optimize both of them in order to obtain the desired output. The tradeoff is presented in figure 1. To avoid high variance, the Bagging Ensemble method is considered very efficient for that aim. This comes from using randomized replication of the original dataset to construct submodules. While the prediction is done through averaging these models' outputs. However, although bagging reduces the variance of one predictor. The bais is still the same taken ordinarily from the original model before the subdivision. Thus, it stays unchanged.

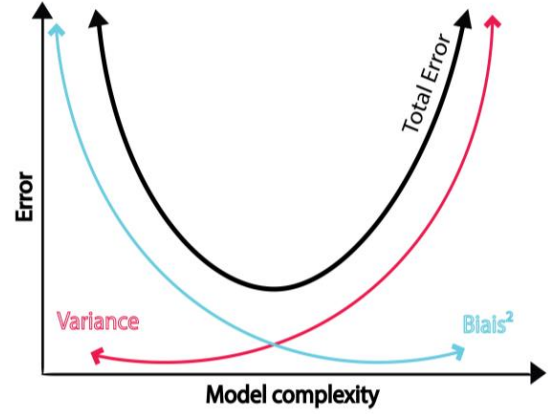

Fig. 1. Bais-variance Tradeoff

## V. DATASET

The Australian government highly courage people to release the transition from traditional resources to renewable energy. It claims to be able of 100% renewables in 2030[5]. The data used for model assessment comes from Alice springs PV plant in Australia. It contains various sensors to follow every slight parameter in the photovoltaic plant. The data collected contains the time indicator, relative humidity, wind speed and orientation, horizontal irradiation, relative horizontal irradiation temperature, and PV power. The database remains is for 3 years from the first of April 2016 to the first of August 2019. This provides sufficient information for training and validation.

## VI. FEATURE ENGINEERING

The relation between PV power over the years makes them a key factor that may contribute to the PV power. From figure 2 and figure 3, the historical PV power is plotted for the month of August and for April from four successive years. The tow plots resume the seasonal variation of photovoltaic energy.

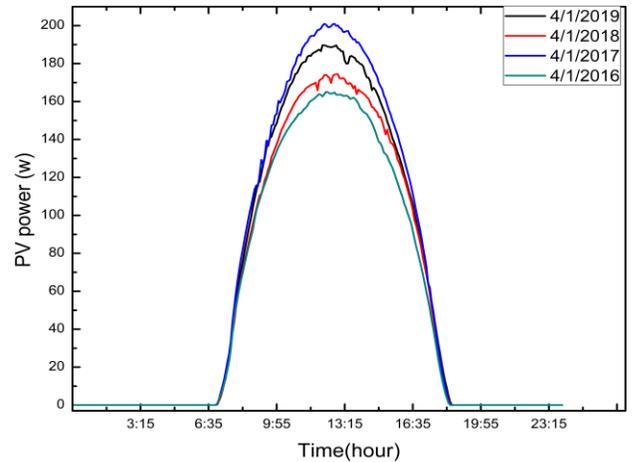

Fig. 2. PV power comparison in the first of April for the previous four years

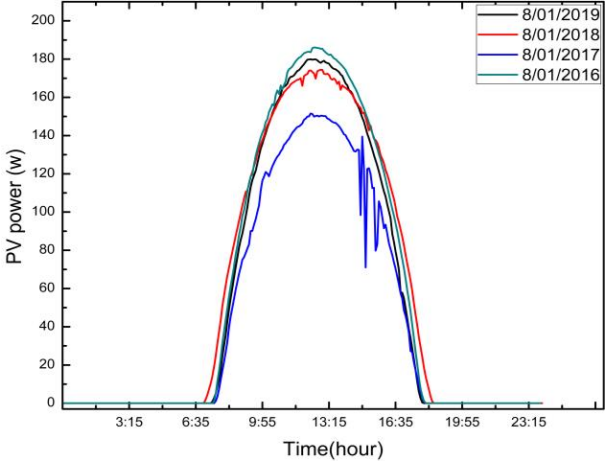

Fig. 3. PV power comparison in the first of August for the previous four years

As we notice in the graphs 2 and 3, the real power generated in 2018 for both seasons is approaching to first previous year same time same day. It means that it can add a general idea about the current PV power. That's why it is added as an important input to estimate the photovoltaic power for the next year 2019. Graph 4 presents the yearly PV power for three successive years.

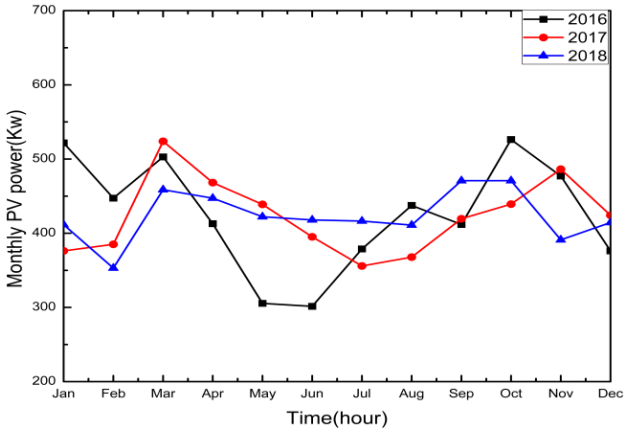

Fig. 4. PV power comparison in the first of April for the previous four years

To sum up, the features used as predictors in our forecasting model are basically weather parameters, PV power generated at the same instant from the previous year and the hourly time indicator. These inputs face many processing stages. The first step is feature engineering and data cleaning. At this level the missing and odd data is deleted. Next the features extracted have weights giving them specific importance basing on their effect in the photovoltaic power generated. The training data is for the last four years from 2016-2019 and the testing data is for two months. The timestep for the data collected is for 5 min. i.e. in one day the data acquired for 288 times.

## VII. FEATURE IMPORTANCE

Variable importance (VI) is taken more attention in statistics and probability. It can be defined as the dependency proportion of the predicted output from the feature's inputs. Important variables or sequence of variables are proportional or inversely proportional to the output to the parameter inputs. Obviously, every domain knowledge has weak indicators and strong indicators in the majority of prediction problems. Feature relative importance (FRI) is a method to classify the input parameters according to the nature of the correlation they have with the output parameters. There are various tools for identifying the importance of each feature. In this paper, we will take a model reliance class (MCR)[34] taken into consideration three feature importance methods: Elastic net, LIME, and XGboost.

### A. Elastic net

Introduced by H. Zou and T. Hastie[35], Elastic Net investigates the relationship between the system features. Given a greater magnitude for the features that follow the behavior of the output function. The said method belongs to linear regression class and it combines two methods Least absolute shrinkage and selection operator (Lasso) and Ridge regressors. This fusion takes advantage of L1 and L2 regularization benefits [35]. The equation behind the model is written in Eq. (5).

$$\frac{\sum_{i=1}^{n}(y_i - x_i^J \beta)^2}{2n} + \lambda(\frac{1-\alpha}{2}\sum_{j=1}^{m}\beta_j^2 + \alpha\sum_{j=1}^{m}|\beta_j|) \quad (5)$$

Where $y_i$ is the output and $x_i$ is the input. Elastic net is chosen in this paper because it outperforms both LASSO and Ridge Regularization for the majority of the cases.

### B. LIME

The second method is Local Interpretable Model-agnostic Explanations (LIME). This method uses a specific representation that imitates the predictor in a meaningful way to identify the importance of the features. Ribeiro et al.[36] uses this approach to explain how the black box works and the correlation between the results and the prediction. The general equation of LIME is presented in Eq. (6)

$$\xi(x) = \arg\min_{g \in G} \mathcal{L}(f, g, \pi_x) + \Omega(g) \quad (6)$$

Where $\mathcal{L}(f, g, \pi_x)$ is who much the function $g$ is approaching to $f$ in a locality $\pi$. with a $g$ is the model explanation, and $\Omega(g)$ is the measure of complexity $\pi$ is the proximity measure between an instance $z$ to $x$, $f(x)$ is a function probability. This method is very efficient in feature selection and widely used in ML algorithms.

### C. XGboost

Extreme Boosting model firstly introduced by Leo Breiman and enhanced by Jerome H. Friedman presents an optimization approach in order to minimize the cost function in regression and classification problems. The said approach builds an efficient single model based on variant weak learners usually taken decision trees. The aforementioned model works as follow, First, a cost function is fixed as a weak hypothesis such as mean square error. And iteratively train the model on how to minimize this error. Taken $F_m$ as a weak forecasting model and assuming that takes the average of the prediction. After training our model the result can be written in Eq. (7).

$$F_{m+1}(x) = F_m(x) + h(x) = y \quad (7)$$

So that

$$h(x) = y - F_m(x) \quad (8)$$

With $h$ is the gradient boosting residual that improves the predictor. This ML algorithm is used in this study to identify the vector feature importance. However, the hyperparameter optimization is not done to the said model giving it the default values of scikit-learn python libraries.

## VIII. FORECASTING MODELS

In this part, some popular ensemble methods are presented to be used in the training and comparison part. These techniques are Bagging, Boosting, extreme boosting and Random forests.

### a) Random decision forests

Bootstrap aggregation given the acronym of Bagging by LEO BREIMAN in 1996[28] consists of merging different predictors which are usually decision trees. This process aims to increase model accuracy. This model covers most of the deficiencies of individual predictors to generate an accurate estimation. Every model is given a specific vote. Then all the votes are combined in order to improve accuracy. This approach can be used in both classification and regression problems. The Equation describes the concept of this method. Assuming we have a database $\{(y_n, X_n), n \in \mathbb{N}\}$ with $x, y$ the numerical input and the output parameters respectfully, and $n$ is the samples number. Predicting $y_k$ is done through the subset $\varphi_k(x, \mathcal{L})$ with $\mathcal{L}$ is the learning set. The key idea is working with multiple predictors and leveraging the outputs $\sum y_k$. This process may lead to an accurate forecast since the variance between the predictors will decrease. Eq. (9) describes more the concept.

$$\varphi_B(x) = E_{B\varphi}(x, \mathcal{L}^{(B)}) \quad (9)$$

where $E$ is the average between the output responses, Obviously, the predictors $\mathcal{L}$ should be different in unstable models. Thus, the results are remarkable and the accuracy is enhanced[15]. Bagging techniques are parallel trained with subsets in order to limit the variance. The result is the mean between all these learners.

Bagging was applied in different areas like medical field [37], energy[38]  such as demand estimation[39]. Regarding Breiman's work, Bootstrap aggregation is efficient in variance reduction. It is frequently used with decision trees (DT) and random forests (RF). Otherwise it is applicable to any machine learning method. One of the best features of this type of ensemble method is avoiding overfitting and reducing variance.

RF proposed by Tin Kam Ho[40] is one of the popular applications of bagging due to its great efficiency with less computational time. It can handle both classification and regression nonlinear tasks. It is literally a large number of DT running at the same time. Thus, the result is the maximum votes from individual ensemble trees. From that perspective the name forest comes as a definition of ensemble trees. RF is firstly used for classification then it is extended to handle regression tasks with numerical values outputs. From that standpoint, the term Classification And Regression Trees (CART) is introduced. RF consists of a map of subsystems trees with their costs if taken in the shape of an ensemble of DT. The variance is used as an indicator of decision impurity. Every internal node in the DT unit indicates the beginning of a binary questionnaire. The answer will split other queries called edges till there is no other detail is not mentioned. In other words, the node cannot be divided anymore. The result of this tree is called a leaf. The regression tree forecasts the continuous values basing on that internal subdivision with the features inputs. The latter technique accuracy is basically relying on the splitting settings and the leaf node ending conditions for each branch. Taking the variance as a significant parameter for measuring the impurity, the leaf is the output of each tree and the prediction is done through the majority of the forecasts. the prediction system and the rooted mean square error is the variance value. The pruning technique is an optimization technique related to DT to eliminate the branches that don't give additional information to the predictor. Thus, the complexity of the system will decrease. Although DT reduces the variance scientifically. The bais remain the same from the first subsystem. In this study, weighting importance is integrated to decrease the aforementioned element.

### B. Gaussian Naïve Bayes

Gaussian Naïve Bayes (GNB) is a simple ML method frequently used in classification tasks but it performs well on regression problems with supervised learning [41]. It consists of modeling the output distribution with kernel density estimators through Bayes theorem. Assuming that $p(Y/E)$ is the probability density of inputs $E$ to the target $Y$. the equation of GNB is written as follow.

$$P(Y|E) = \frac{p(E,Y)}{\int p(E,Y)dY} = \frac{p(E|Y)p(Y)}{\int p(E|Y)p(Y)dY} \quad (10)$$

With $p(X_i, Y)$ is predicted separately.

### C. K-Nearest Neighbors

K-nearest neighbors (KNN) is a statistical approach. The philosophy behind this approach is selecting $k$ samples closer to unknown labels. The latter is calculated by averaging the labels of the features near to the unknown label. The Euclidean distance is often implemented through KNN to measure the distance between two objects in Eq. (11).

$$D(a,b) = \sqrt{\sum_{i=1}^{n}(x_i - y_i)^2} \quad (11)$$

It should be mentioned that the suitable value of $k$ is mandatory to provide optimum accuracy while using this ML method. Therefore, in our case a grid search is used to tune KNN.

*D. Proposed approach*

The proposed method consists of including voted weights vector into the ML black box. Then, the said feature importance vector is fed to a RF, multimodal to generate an accurate PV power output. Assuming we have $n$ features, the prediction system is divided into $n$ subsystems. For each subsystem, a $k$ feature parameter is eliminated from the database. With Bagging model, every subsystem gives a prediction output $y_i$. Let's be $w_i \in [0,1]$ [d] the rate of importance of each feature. The final output is concluded through summing the weighted subsystems products by an importance factor. Eq (12) will explain the process.

$$Y_i = \sum_{i=1}^{n} w_i \overline{y_i} \quad (12)$$

In this vein, the weight values $\alpha_i$ are adjusted using three successful methods with FRI method. These methods are LIME [42], Elastic Net and extreme boosting. The usefulness of using these three techniques together comes from the variant architectures between these tools. In order to overcome the problem of which importance vector is going to be used, these three methods are taken into consideration. Assuming N is the number feature weighted tools. The relation between them can be averaging or voting. Eq. (13) present the process.

$$w_{feature} = \frac{1}{N}\sum_{j=1}^{N} w_j \quad (13)$$

In this paper averaging is the primarily case used. Then, we try to use different voting percentages between the methods. The proposed method allows Random forests to overcome overfitting and add more robustness to the model. Taken from the literature, Kolcz and Teo[43] proves that feature weighting is able to improve significantly the predictor robustness with less computational work. By using differently feature importance, we claim that using multiple techniques can outperform one technique in Eq.14. The idea of importance vector comes to add a feature reweighting coefficient $\overline{w}$. For the sake of explanation, $\bar{x}_{ij}$ is the feature weighted of $x_i$ which is calculated through Eq.14.

$$\overline{x_{ij}} = x_{ij} / s(w_j), \quad j=1,\ldots,d \quad (14)$$

With $s$ is a positive coefficient, $w$ is the weight vector and $x_{ij}$ is the j$^{th}$ feature of $x_i$. The algorithm for the proposed approach is presented in Table 1

Table 1. Proposed algorithm

**Algorithm: Hybrid model**

**Input :** Data acquisition.
  Step 1: Feature importance determination using LIME, ELASTIC Net, XGBoost.
  Step 2: Vector importance $I = \{I_1, I_2, \ldots, I_n\}$ according to averaging or voting.
  Step 3: Creating an ensemble of databases $n$. Each database eliminates one feature $x_i$.
  Step4: calculate multiple predictions $y_i$ using each database with ensemble model.

**Output:** The result is the sum of each prediction $y_i$ multiplied by the covariance $w_i \in [0,1]$ of the feature importance missed.

IX. EXPERIMENTAL RESULTS

The data collected is split into 80% for model training and 20% for testing and evaluation. The hybrid model is tuned using a randomized search. The hyperparameters list includes Max depth, Min sample leaf, and the Max leaf nodes. It had been determined through Randomized Search tool.

Table 2. Table errors

| Hyper parameters | Value |
|---|---|
| Random state | 0 |
| Min sample leaf | 20 |
| Max leaf nodes | 100 |
| Max depth | 8 |

The model is simulated in a Lenovo laptop Lenovo Ideapad 720S-15IKB (i7 with 8 CPU cores) and the paralyzed processing provided by the NVIDIA GPU package is integrated to accelerate the model training and testing. Regarding the said parameters, the model is training and cross-validated with ten folds. Then the training model is tested to evaluate its performance. For an assessment purpose, score metrics calculated are the mean square error, mean absolute error and median absolute percentage error.

$$RMSE = \sqrt{\frac{1}{n}\left(\sum_{i=1}^{n}(\bar{y}_i - y_i)^2\right)} \quad (15)$$

$$MAE = \frac{1}{n}\sum_{i=1}^{n}(\bar{y}_i - y_i) \quad (16)$$

*A. Database and forecasting horizon*

In the simulation part, a yearly dataset from 2017 to mid-2019 is used for training and the month of August 2019 is for the evaluation process. The Australian database used in our experimental analysis provides rich information with a time step of 5 minutes. The weather database from Alice springs provides the temperature, the relative humidity, the wind speed and direction, the PV power and the horizontal and vertical radiation. Figure 6 presents the quantity of samples used for training and testing.

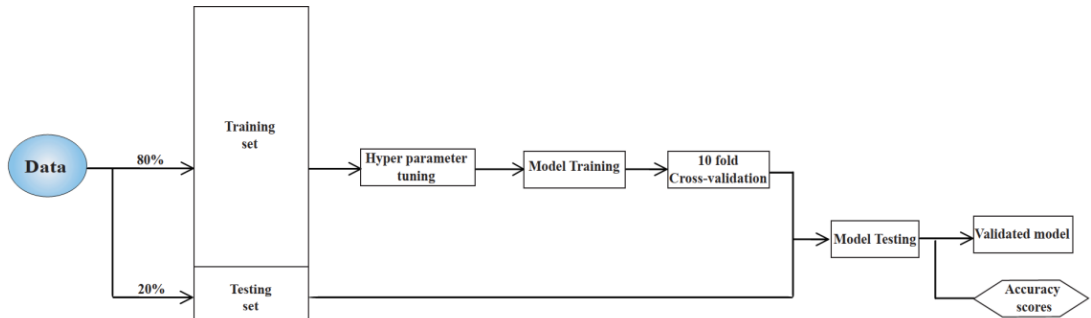

Fig. 5. Validation procedure

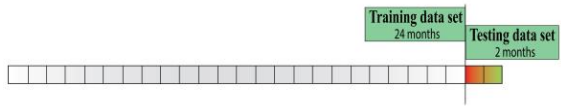

Fig. 6. Forecasting Horizon

*B. Feature importance*

The results from these methods shape the vector importance metrics. Horizontal power and previous PV power from the same instance in the neighboring year had much importance as well as the previous PV power. The other features are poorly influencing the system. Figure 7 presents the results of feature importance using LIME in blue color, Elastic Net with the green color and Extreme boosting in red color.

The method is done on bagging techniques to adjust high variant bais and in our example. The cross-validation results of the simulation are shown in figure 8.

*C. Simulation results*

The simulation results are done using python coding and compared with various ML models namely KNN, Gaussian naïve Bays and Decision Trees. The said models use literally the same conditions to ensure a fair comparison including using a randomized search for hyperparameters tuning and the previous dataset. The simulation targets the forecasting of summer months namely June and July 2019 referring to two years (2017-2018). The database for training is taken from an Australian PV plant. The testing data is evaluated using two popular score metrics: the RMSE and the MAE. Figure 9,10 and 11 presents the shape of the forecasting results from 2 months, one week and one day respectively.

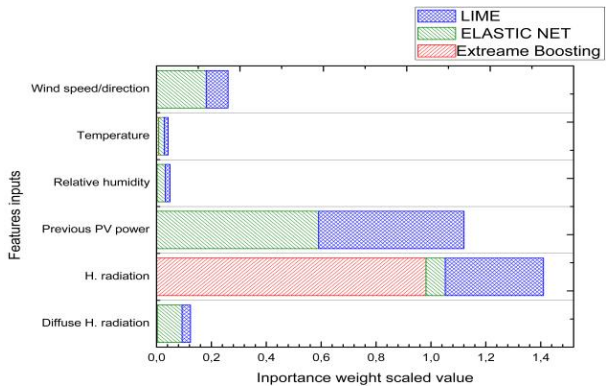

Fig. 7. Relative influence of the eleven input variables on the target variation for the Australian weather database.

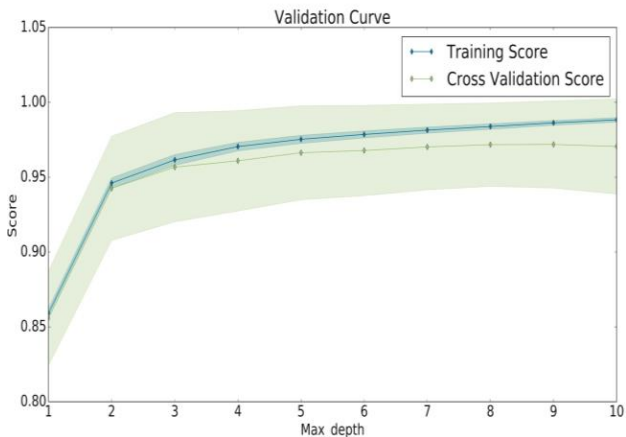

Fig. 8. Cross validation graph in terms of R²

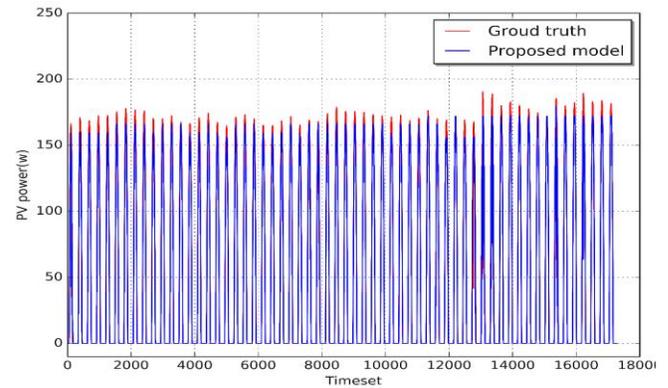

Fig. 9. June-July PV power (w) forecasting using the proposed method

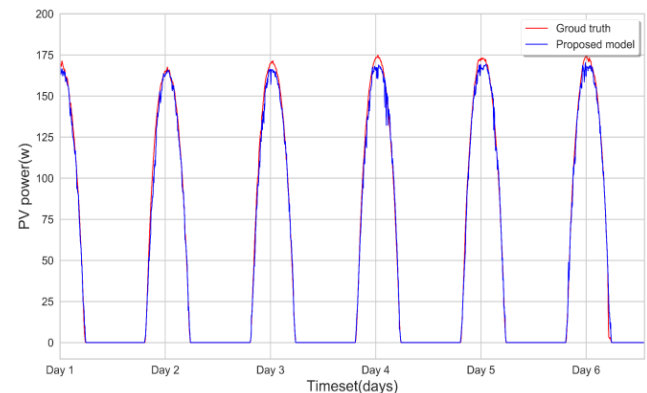

Fig. 10. Week-Ahead PV power (w) forecasting proposed method

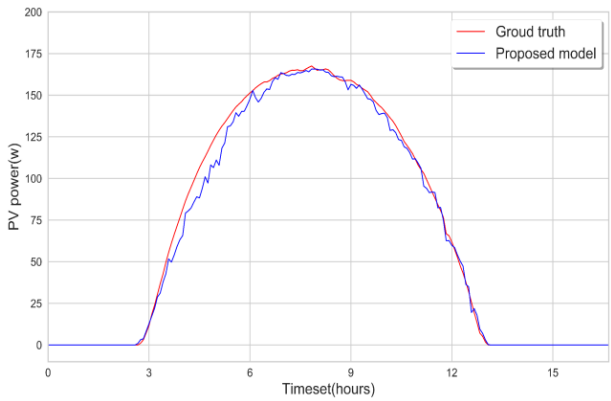

Fig. 11. Day-Ahead PV power (w) in August 2019.

Regarding the graphs, it has been noticed that the model proposed provides a great precision referring to the points matched between the ground truth and the forecasted PV power. The real PV power in red is slightly different than the predicted power in mid-day. i.e. when the photovoltaic power is generated at its maximum. PV power predicted is following the real PV power.

For the sake of comparison, the proposed approach is simulated with the state-of-the-art forecasting models namely KNN, Gaussian Naïve Bays and the original Decision trees. The error values are calculated and shown in figure 14 and table 1.

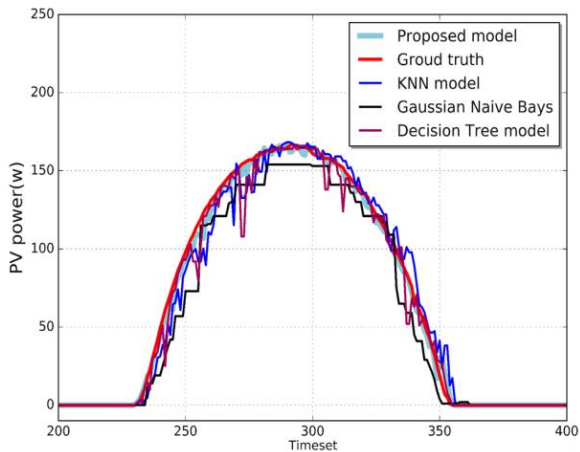

Fig. 12. PV power forecasting methods comparison

Table 4. Table errors

| Models | RMSE | MAE |
|---|---|---|
| DT | 9.88 | 3.46 |
| KNN | 7.35 | 4.80 |
| Gaussian NB | 13.59 | 7.45 |
| Proposed Method | 6.49 | 2.61 |

It has been noticed that the proposed method outperforms all the forecasting models. This accuracy comes from an improvement in terms of bais. Through the feature importance vector. Nevertheless, the preprocessing is heavy and time consuming regarding the weight extraction. Nevertheless, the proposed model is suitable in time series forecasting.

The accuracy is high which proves the robustness of the model built. For the aim of testing the model in the seasonal change. Four days from different seasons are simulated and plotted in graph 14.

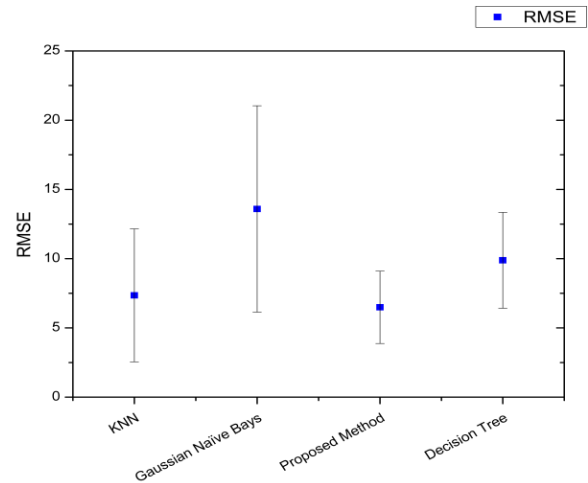

Fig. 13. RMSE/MAE models comparison

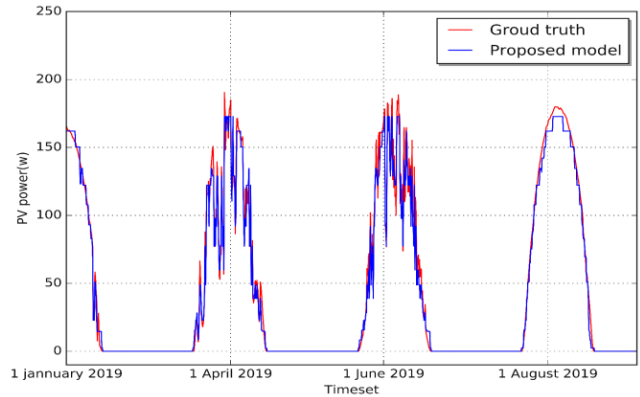

Fig. 14. Results of forecasting models

To get deeper to the partition vector importance, three cases are studied, the first consists of an equal partition of the vector importance between them then 3 various cases are presented in table 4 and the accuracy of each case is presented in terms of RMSE and MAE. The error doesn't change too much so the proposed method. However, the accuracy is high in terms of months which proves the robustness of the model built.

Table 3. Table errors

| Method/Error | Average Proportion | Case 1 | Case 2 | Case 3 |
|---|---|---|---|---|
| LIME | 1/3 | 50% | 20% | 50% |
| ELASTIC NET | 1/3 | 30% | 30% | 20% |
| XGBOOST | 1/3 | 20% | 50% | 30% |
| RMSE | 9.45 | 7.13 | 7.19 | 7.4 |
| MAE | 3.64 | 3.5 | 3.4 | 3.7 |

*D. Discussion*

From Fig. 7, the relative influence of the eleven variables on the target variation gives a different result. knowledge importance differs from one technique to another. XGboost gives a higher error value for Horizontal radiation while ELASTIC Net proves that the forecasted power relies on both previous PV power and the wind effect with the latter being less effective. on the previous PV power and with a lower value the wind effect. On the contrary, LIME gives the most expected results. Thus, the previous PV power and H. radiation are the most correlated features with the current PV power. Each feature is given a relevant importance value $w_i$ with $\sum_{i=1}^{n} w_i = 1$. The weights are calculated based on Chernoff-Hoeffding Theory[44]. Another approach is based on the calculation the weights by taking into consideration the errors from the features importance methods while shifting and deducing the percentage of error caused by the feature missed. This feature ranking gives a remarkable result. The error is low taken from 2 months of forecasting. Cross-validation is applied to the proposed method and gives a 96% of accuracy in terms of squared learning rate. This is done by testing the enhanced RF. This proves the robustness of the proposed method. It can be concluded that relying on domain knowledge is contributing significantly to the prediction accuracy. The major advantage of the proposed procedure is that the random forest units prevent the system from the individual error caused by a unique feature. Although the preprocessing is low and takes time to establish the domain knowledge, the results are worthy. Taking as example a system with a high dimension, when feature selection is not sufficient to reduce the noise. The proposed technique is efficient in RF and diminishes the errors coming from the misleading inputs. A fair comparison of the enhanced RF with KNN, Gaussian NB, and DT shows that the proposed technique is outperforming the aforementioned techniques. The high variability of weather conditions from four seasons shows that the proposed model is staying following to the ground truth. From figure 14, the predicted power follows the real values with a small error caused by the sudden disturbance of inputs parameters. Finally, a variety of error percentage is done through changing the percentage of each feature importance methods influence the importance vector error. A small change has been noticed due to the small number of features used in PV power forecasting. More investigation on a high dimensional system is needed to show the relative contribution of the proposed method for forecasting accuracy.

## X. CONCLUSION

This paper discusses a new approach relaying on feature importance in ML models. Indeed, FI not only shrinks the input parameters and eliminates the misleading features, but also it interferes with the predator. RF as a bagging ensemble method is used as an application for this approach to forecasting photovoltaic power during two months. The importance vector is calculated with three popular methods namely LIME, XGboost, and Elastic Net. These methods enrich the system by giving different outputs. With the use of domain knowledge and the elimination of one feature for each simulation. The combination of feature importance and ensemble methods contribute to the prediction accuracy. The proposed method enhanced significantly the forecast results making it extremely suitable for time series forecasting. The obtained RMSE from this method is 6.49 whereas the MAE is 2.46. Thus, the proposed method outperforms all the benchmarked methods including K nearest neighbors, Gaussian naïve bays and Decision trees. The domain knowledge enhanced prediction accuracy. forecast the medium-term forecasting aid in preventive maintenance scheduling and investment planning. Hence, the drawbacks of this method can be resumed essentially to two factors:

- A clear vision of the output behavior is required.
- The correlated features can be harmful to model accuracy. Therefore, a serious dimensionality reduction is needed to reduce this effect.

More investigation on the implementation of feature importance on the rest of ML algorithms is planned as a future work of this study.

## ACKNOWLEDGMENTS

The authors would highly acknowledge the financial support of the Qatar National Research Fund (a member of Qatar Foundation). Also special thanks to Pr. Haithem Abu-Rub from Smart Grid Center Laboratory SGC who made this study possible.

## REFERENCES

[1] A. Aleman et al., "Perceived Reality and Loudness of Auditory Verbal Hallucinations Is Associated With Reduced Connectivity Between Thalamus and Auditory Cortex : an Fmri Study in Schizophrenia Patients," *Schizophr. Bull.*, vol. 35, no. 2003, pp. 178–178, 2009.

[2] N. L. Panwar, S. C. Kaushik, and S. Kothari, "Role of renewable energy sources in environmental protection: A review," *Renew. Sustain. Energy Rev.*, vol. 15, no. 3, pp. 1513–1524, 2011.

[3] E. Ogliari, A. Dolara, G. Manzolini, and S. Leva, "Physical and hybrid methods comparison for the day ahead PV output power forecast," *Renew. Energy*, vol. 113, pp. 11–21, 2017.

[4] D. Graupe, D. J. Krause, J. B. Moore, and J. B. Moore, "Identification of Autoregressive Moving-Average Parameters of Time Series," *IEEE Trans. Automat. Contr.*, vol. 20, no. 1, pp. 104–107, 1975.

[5] T. Lin, B. G. Horne, P. Tiiio, and C. L. Giles, "Learning Long-Term Dependencies in," *IEEE Trans. Neural Networks*, vol. 7, no. 6, pp. 1329–1338, 1996.

[6] A. Dolara, S. Leva, and G. Manzolini, "Comparison of different physical models for PV power output prediction," *Sol. Energy*, vol. 119, pp. 83–99, 2015.

[7] M. G. De Giorgi, P. M. Congedo, and M. Malvoni, "Photovoltaic power forecasting using statistical methods: Impact of weather data," *IET Sci. Meas. Technol.*, vol. 8, no. 3, pp. 90–97, 2014.

[8] A. Yona, T. Senjyu, T. Funabshi, and H. Sekine, "Application of neural network to 24-hours-ahead generating power forecasting for PV system," *IEEJ Trans. Power Energy*, vol. 128, no. 1, 2008.

[9] M. Abdel-Nasser and K. Mahmoud, "Accurate photovoltaic power forecasting models using deep LSTM-RNN," *Neural Comput. Appl.*, vol. 31, no. 7, pp. 1–14, 2017.

[10] A. Alzahrani, P. Shamsi, C. Dagli, and M. Ferdowsi, "Solar Irradiance Forecasting Using Deep Neural Networks," *Procedia Comput. Sci.*, vol. 114, pp. 304–313, 2017.

[11] L. P. Kaelbling, M. L. Littman, and A. W. Moore, "Kaebling & Moore (1996) Reinforcement learning - a survey," pp. 1–49, 1997.

[12] *Ensemble Machine Learning*. 2012.

[13] R. K. Agrawal, F. Muchahary, and M. M. Tripathi, "Ensemble of relevance vector machines and boosted trees for electricity price forecasting," *Appl. Energy*, vol. 250, no. May, pp. 540–548, 2019.

[14] R. E. Schapire, "A brief introduction to boosting," *IJCAI Int. Jt. Conf. Artif. Intell.*, vol. 2, pp. 1401–1406, 1999.

[15] L. Breiman, "Bagging predictors: Technical Report No. 421," *Dep. Stat. Univ. Calif.*, no. 2, p. 19, 1994.

[16] M. Rana, I. Koprinska, and V. G. Agelidis, "Univariate and multivariate methods for very short-term solar photovoltaic power forecasting," *Energy Convers. Manag.*, vol. 121, pp. 380–390, 2016.

[17] T. Cai, S. Duan, and C. Chen, "Forecasting power output for grid-connected photovoltaic power system without using solar radiation measurement," *2nd Int. Symp. Power Electron. Distrib. Gener. Syst. PEDG 2010*, pp. 773–777, 2010.

[18] E. Lorenz, J. Hurka, D. Heinemann, and H. G. Beyer, "Irradiance Forecasting for the Power Prediction of Grid-Connected Photovoltaic Systems," *IEEE J. Sel. Top. Appl. Earth Obs. Remote Sens.*, vol. 2, no. 1, pp. 2–10, 2009.

[19] R. Al Iqbal, "Empirical learning aided by weak domain knowledge in the form of feature importance," *Proc. - 2011 Int. Conf. Multimed. Signal Process. C. 2011*, vol. 1, pp. 126–130, 2011.

[20] L. A. Fernandez-Jimenez *et al.*, "Short-term power forecasting system for photovoltaic plants," *Renew. Energy*, vol. 44, pp. 311–317, 2012.

[21] D. Thevenard and S. Pelland, "Estimating the uncertainty in long-term photovoltaic yield predictions," *Sol. Energy*, vol. 91, pp. 432–445, 2013.

[22] T. K. Bhattacharya and T. K. Basu, "Medium range forecasting of power system load using modified Kalman filter and Walsh transform," *Int. J. Electr. Power Energy Syst.*, vol. 15, no. 2, pp. 109–115, 1993.

[23] U. K. Das *et al.*, "Forecasting of photovoltaic power generation and model optimization: A review," *Renew. Sustain. Energy Rev.*, vol. 81, no. August 2017, pp. 912–928, 2018.

[24] A. Meunkaewjinda, P. Kumsawat, K. Attakitmongcol, and A. Srikaew, "Grape leaf disease detection from color imagery using hybrid intelligent system," *5th Int. Conf. Electr. Eng. Comput. Telecommun. Inf. Technol. ECTI-CON 2008*, vol. 1, pp. 513–516, 2008.

[25] Z. Wan, G. Wang, and B. Sun, "A hybrid intelligent algorithm by combining particle swarm optimization with chaos searching technique for solving nonlinear bilevel programming problems," *Swarm Evol. Comput.*, vol. 8, pp. 26–32, 2013.

[26] J. Gao and B. Liu, "Fuzzy multilevel programming with a hybrid intelligent algorithm," *Comput. Math. with Appl.*, vol. 49, no. 9–10, pp. 1539–1548, 2005.

[27] S. B. Kotsiantis, "Decision trees: A recent overview," *Artif. Intell. Rev.*, vol. 39, no. 4, pp. 261–283, 2013.

[28] L. Breiman, "Bagging predictions," *Mach. Learn.*, vol. 24, no. 2, pp. 123–140, 1996.

[29] M. Kearns, "Thoughts on hypothesis boosting," *Unpubl. Manuscr.*, vol. 45, p. 105, 1988.

[30] P. Kazienko, E. Lughofer, and B. Trawiński, "Hybrid and ensemble methods in machine learning," *J. Univers. Comput. Sci.*, vol. 19, no. 4, pp. 457–461, 2013.

[31] J. M. Corchado and J. Aiken, "Hybrid artificial intelligence methods in oceanographic forecast models," *IEEE Trans. Syst. Man Cybern. Part C Appl. Rev.*, vol. 32, no. 4, pp. 307–313, 2002.

[32] S. Geman, E. Bienenstock, and R. Doursat, "Neural Networks and the Bias/Variance Dilemma," *Neural Comput.*, vol. 4, no. 1, pp. 1–58, Jan. 1992.

[33] A. Kagan and L. A. Shepp, "Why the variance?," *Stat. Probab. Lett.*, vol. 38, no. 4, pp. 329–333, Jul. 1998.

[34] A. Fisher, C. Rudin, and F. Dominici, "All Models are Wrong but many are Useful: Variable Importance for Black-Box, Proprietary, or Misspecified Prediction Models, using Model Class Reliance," no. Vi, 2018.

[35] H. Zou and T. Hastie, "Regularization and variable selection via the elastic net," *J. R. Stat. Soc. Ser. B Stat. Methodol.*, vol. 67, no. 2, pp. 301–320, 2005.

[36] M. T. Ribeiro, S. Singh, and C. Guestrin, "'Why should i trust you?' Explaining the predictions of any classifier," *Proc. ACM SIGKDD Int. Conf. Knowl. Discov. Data Min.*, vol. 13-17-Augu, pp. 1135–1144, 2016.

[37] M. C. Tu, D. Shin, and D. K. Shin, "Effective diagnosis of heart disease through bagging approach," *Proc. 2009 2nd Int. Conf. Biomed. Eng. Informatics, BMEI 2009*, pp. 1–4, 2009.

[38] E. M. de Oliveira and F. L. Cyrino Oliveira, "Forecasting mid-long term electric energy consumption through bagging ARIMA and exponential smoothing methods," *Energy*, vol. 144, pp. 776–788, 2018.

[39] B. Patrick, D. Nekipelov, S. P. Ryan, and M. Yang, "American Economic Association Machine Learning Methods for Demand Estimation Author ( s ): Patrick Bajari , Denis Nekipelov , Stephen P . Ryan and Miaoyu Yang Source : The American Economic Review , Vol . 105 , No . 5 , PAPERS AND PROCEEDINGS OF THE One H," vol. 105, no. 5, 2018.

[40] T. Kam Ho, "Random Decision Forests Perceptron training," *AT&T Bell Lab.*, 1995.

[41] E. Frank, L. Trigg, G. Holmes, and I. H. Witten, "Technical note: Naive Bayes for regression," *Mach. Learn.*, vol. 41, no. 1, pp. 5–25, 2000.

[42] R. Guo and Wenhui Wu and R. Guo and Wenhui Wu, "Progress in Graphene-Based Electrochemical Biosensors for Cancer DiagnosisProgress in Graphene-Based Electrochemical Biosensors for Cancer Diagnosis," *Gen. Chem. Chem. Chem.*, vol. 2, no. 3, pp. 89–96, 2016.

[43] A. Kołcz and C. H. Teo, "Feature Weighting for Improved Classifier Robustness," *Sixth Conf. Email AntiSpam CEAS*, no. 1, p. 2009, 2009.

[44] A. Siegel and A. Srinivasan, "Schmidt, srinivasan," vol. 8, no. 2, pp. 223–250, 1995.